\documentclass[prl,aps,twocolumn]{revtex4}
\begin{document}

\draft      
\title{Phonon Bloch oscillations in acoustic-cavity structures}
\author{N. D. Lanzillotti Kimura$^1$, A. Fainstein$^1$, and B. Jusserand$^2$}     
\address{$^1$Centro At\'omico Bariloche \& Instituto Balseiro, C.N.E.A., 
8400 S. C. de Bariloche, R. N., Argentina}       
\address{$^2$Laboratoire de Photonique et de Nanostructures, CNRS, Route de Nozay, 
91460 Marcoussis, France}
                                                                                                                                       
\begin{abstract}                                                                
We describe a semiconductor multilayer structure based in
acoustic phonon cavities and achievable with
MBE technology, designed to display acoustic
phonon Bloch oscillations.  We show
that forward and backscattering Raman spectra give
a direct measure of the created phononic Wannier-Stark ladder. We also
discuss the use of femtosecond laser impulsions for the generation 
and direct probe of the induced phonon Bloch oscillations. We propose
a gedanken experiment based in an integrated phonon 
source-structure-detector device, and we present calculations of 
pump and probe time dependent optical 
reflectivity that evidence temporal beatings in agreement 
with the Wannier-Stark ladder energy splitting.

\end{abstract}    

\maketitle   

Electronic Bloch oscillations, that is, oscillations of an 
electron induced by a {\em constant} electric field
in the precense of a periodic potential,~\cite{Bloch} are a 
beautiful and clear
example of quantum effects in solids. When an electric field
is applied on a charged particle in a crystal, 
its wavevector increases with time. Thereafter, Bragg interference
leads to a velocity reduction, and finally to a sign 
change at the band edge. Notwithstanding its simplicity, for 
many years the issue was
controversial and only quite recently the existence of electronic 
Bloch oscillations  has been definitively established. 
In normal metals the large Brillouin zone and electron relaxation 
lead to an overdamped behavior characterized by Ohm's
law. Instead, electronic Bloch oscillations are observable in 
SL's due to the Brillouin zone reduction.~\cite{SL's}

Very recently photon Bloch oscillation devices based in 
optical microcavities have been proposed,~\cite{Malpuech} and
related reflectivity and time-resolved 
optical transmission features have been observed in structures
grown on porous silicon.~\cite{PRL's} The concept behind these devices
is also quite nice and simple. The optical microcavities provide the 
discrete photonic states which can couple through photon mirrors thus 
leading to photonic minibands. The spatial dependent energy gradient, 
equivalent to the electric field for electrons, can be
achieved, e.g., by varying the refractive index.~\cite{Bloch}
As compared to electrons, photon dephasing mechanisms are less effective
and thus Bloch oscillations are, in principle, more easy to observe.
On the other hand, photonic minibands require a large number of optical 
microcavities (around 30-50), 
amounting to total thicknesses larger than 40-50 microns.~\cite{Malpuech} 
This is too large for current molecular beam epitaxy (MBE) 
semiconductor technology.
For this reason the reported structures were grown using electrochemical
methods of porous silicon nanostructuring that do not have this 
limitation.~\cite{PRL's} 
The drawback is that the quality of the samples is not as good, layer interfaces
and refractive indices are not that well controlled, and optical residual 
absorption becomes an important issue in the performance of the structures.

Sound in solid media is also described by a wave equation, the
relevant parameters being the material density and sound velocity,
and the boundary conditions establishing the connection 
between displacement and strain between two different materials.
Extending the above concepts applied to electrons and photons
to phononic structures, in this letter we propose semiconductor 
multilayer structures capable of displaying {\em acoustic phonon} Bloch  
oscillations. The building block of the structures are acoustic phonon 
cavities as recently introduced by Trigo and coworkers.~\cite{Trigo}.
The phonon wavelength is only a few nanometers (phonon 
frequencies being in the THz range), and thus the sample full size and 
required interface layer quality are 
achievable with actual MBE technology. In addition, acoustic phonon
mean free paths are large~\cite{Philosophical} compared to the
structure size (typically below a micron) and thus dephasing is not
a critical issue. We present calculations of phonon reflectivity, displacement
distribution, and time evolution upon excitation with a localized
phonon source to illustrate the device behavior and to demonstrate
the existence of a phonon Wannier-Stark ladder (WSL) and Bloch oscillations (BO).
In addition, using a photoelastic model we calculate the Raman 
spectra which we show give  a direct measure of the created phononic WSL. 
Phonon BO can be independently probed 
using coherent phonon generation techniques. 
For this purpose we propose a specifically designed device
and we present calculations of time dependent reflectivity induced
by femtosecond impulsions in this structure.

A periodic stack of two materials with contrasting acoustic impedances
reflects sound.~\cite{Narayanamurti}  The first $k=0$ folded phonon 
mini-gap in a SL is maximum with the layer´s thickness 
($d$) ratio given by $d_1/v_1=d_2/3v_2$, 
the stop-band and reflectivity of such a phonon mirror being 
determined by the acoustic impedance mismatch 
$Z=\rho_1v_1/\rho_2v_2$ and the number of SL 
periods.~\cite{Lacharmoise,Jusserand} 
A phonon cavity can be constructed by enclosing
between two SL's a spacer of thickness 
$d_c=m\lambda_{c}/2$, where
$\lambda_{c}$ is the acoustical phonon wavelength at
the center of the phonon minigap.~\cite{Trigo}  
The cavity confined modes correspond to discrete energy
states within the phonon stop-band, their width (i.e., the cavity
Q-factor) being determined by the phonon mirror reflectivity.~\cite{Lacharmoise}
When a large series of phonon microcavities are coupled 
one after the other, the discrete confined energy states form
phonon bands that resemble the minibands in
electronic SL's. In order to display Bloch oscillations,
the energy of the $i$-th cavity must differ from that of the $(i-1)$-th
in a constant value. Such a linear dependence
with position of the phonon cavity-mode energy, analogous to an 
electric field for electrons, can be obtained by tuning the cavity
widths.~\cite{Malpuech} We thus consider a multilayer structure
where each unite cell consists of an acoustic phonon mirror made by $(n+1/2)$ 
$\frac{\lambda}{4}/\frac{3\lambda}{4}$ periods of two materials with contrasting 
acoustic impedances (GaAs/AlAs in the examples discussed here), followed 
by a $\lambda$ cavity (GaAs). This unit cell is repeated
$N_{c}$ times, with layer thicknesses increasing from the
surface to the substrate so as to have a linear decrease of cavity-mode 
energy by steps of $\Delta$.

The stationary solutions of the acoustic waves in the proposed 
structure can be derived using a matrix method implementation 
of the elastic continuum model.~\cite{Trigo}
The calculations give $(i)$ the phonon field distribution in the 
different layers, $(ii)$ the phonon reflectivity and/or transmission, and
$(iii)$ the variation along the structure of the energy bands
associated to an infinite SL with the local unit cell. 
The latter are given by the condition $-1 \leq (a_{11} + a_{22})/2 \leq 1$, 
where $a_{11}$ and  $a_{22}$ are the diagonal elements of the transfer
matrix across each period of the structure.~\cite{Malpuech} 
In Fig.~\ref{fig1}
we present results for a $N_{c}=25$ period acoustic cavity structure. 
The first unit cell is made by a 2.5 GaAs/AlAs period $59.3\AA/23.5\AA$
phonon mirror ($n=2$) and a GaAs spacer tuned to 20~cm$^{-1}$ 
($79\AA$). The energy steps are given by $\Delta=0.15$~cm$^{-1}$,
and the structure is limited from the two sides by GaAs. Panels
$(a)$ to $(c)$ display, respectively, the phonon band structure 
(black regions represent ``forbidden'' energies),
the phonon reflectivity, and the phonon displacement distribution
as a function of position and energy. In the latter panel,
calculated for phonons entering from the left, darker regions indicate 
larger acoustic phonon intensities. Several features
should be highlighted in these figures. $(i)$ An acoustic phonon
allowed band originated in the coupled discrete cavity modes is 
observed between two forbidden minigap bands (panel (a)). The 
energy of the bands decreases linearly with cavity number 
according to design. Three different spectral regions can be 
identified. Between 17.4 and 18.6 cm$^{-1}$ (shaded in Fig.~\ref{fig1}),
phonons are confined in a spatial region limited by the top and bottom 
of the lower and upper minigap bands, respectively. On the other hand,
below(above) 17.4(18.6)~cm$^{-1}$ a phonon entering from cavity $\#1$ will 
bounce back at the bottom of the lower (upper) minigap band 
and leave the sample. $(ii)$ The phonon reflectivity (panel (b)) 
displays a broad stop-band (between $\sim 15$ and $\sim 21.5$ 
cm$^{-1}$), basically determined by the superposition
of the individual cavity minigaps. A series of dips modulate 
the reflectivity within this wide stop-band.
Between the lower stop-band limit  
and $\sim 17.4$~cm$^{-1}$, and between $\sim 18.6$~cm$^{-1}$
and the upper stop-band limit, a series of 
features with varying energy spacing can be identified. These originate
from phonon interferences determined by a propagation limited 
by the sample surface and by the minigap bands.
These features are relatively weak and broad because of the 
small acoustic impedance mismatch between the top GaAs layer 
and the cavity structure.
On the other hand, sharp reflection dips are observed in the
region between $\sim 17.4$ and $\sim 18.6$~cm$^{-1}$. These dips, which are 
equidistant and separated by $\Delta=0.15$~cm$^{-1}$ correspond to the 
coupling of external phonons,
by tunneling through the minigap band, to the Wannier-Stark states confined 
within the structure. $(iii)$ This phononic WSL can
be also identified by the well defined discrete phonon modes 
displayed in panel (c). The WSL is the spectral domain counterpart of 
the BO. It is precisely in this spectral region where
oscillations should appear in the time-domain.

The time and spatial variation of the acoustic phonon displacement
field $U_g(z,t)$, created by a pulse described by a spectral 
function $g(\omega)$ and incident at $t=0$ at the GaAs-sample 
interface ($z=0$), can be evaluated using the scattering method 
described by Malpuech and Kavokin.~\cite{Malpuech} Within this 
description, $U_g(z,t)=\frac{1}{2\pi}\int_{-\infty}^\infty 
u(z)g(\omega)exp(-i\omega t)d\omega$, where $u(z)$
are the stationary solutions of the elastic wave equation with
frequency $\omega$ shown in Fig.~\ref{fig1}(c). 
The time evolution of such wavepackets
for the two different energies indicated in Fig.~\ref{fig1}(a)
are shown in Fig.~\ref{fig2}. For energies above or below the
phonon WSL region (20 cm~$^{-1}$ in the example shown), 
the incident pulse propagates within 
the sample up to a position where it is back-reflected by
a mini-gap band leaving afterwards the sample. On the contrary,
when $\omega$ corresponds to the WSL energy region 
(18.3 cm~$^{-1}$ in the displayed figure), a fraction of 
the pulse energy is backreflected at the surface while another 
part enters the structure by tunneling through the minigap 
band developing afterwards clear periodic oscillations within the 
structure. In order for these phonon BO
to be observed, the FWHM of $g(\omega)$ should
be larger than $\Delta$. For these calculations
we have used a gaussian distribution with $2\sigma=1.0$~cm$^{-1}$. 
On the other hand, the period of the oscillations (and consequently 
the length travelled by the pulse) is directly determined by $\Delta$ 
($\tau_B=h/\Delta$).

In what follows we will show how Raman scattering 
and coherent phonon generation experiments can provide a direct
probe of the phonon WSL and BO, 
respectively, in these devices. Raman scattering has extensively 
shown to be a powerful tool to study phonons in semiconductor
multilayers~\cite{Jusserand} and, in particular, confined
modes in acoustic cavities.~\cite{Trigo,Lacharmoise}
In order to evaluate the Raman spectra, we use
a photoelastic model.~\cite{Trigo,Jusserand} 
We analyze two experimental geometries, namely
backscattering (BS, $k_S \sim -k_L$, $q=\sim 2k_L$) and forward 
scattering (FS, $k_S \sim k_L$, $q \sim 0$). Here $k$ refers 
either to the scattered or laser wavevector, and $q$ is the 
transferred wavenumber. In Fig.3 we present BS and FS 
spectra calculated for the sample described
above and assuming laser excitation with 550nm. 
For comparison purposes the corresponding phonon reflectivity 
is also shown. The Raman spectra 
display a complex series of peaks in the stop-band spectral
region. We have verified that such rich spectra are a kind
of sample finger-print that can be
used as a characterization tool. Interestingly, the BS and FS
spectra display clear peaks and dips, respectively, at exactly
the WSL energies. 
High resolution Raman set-ups working in the visible can discern
spectral features with resolution better than 
0.02~cm$^{-1}$,~\cite{Pinan} thus providing a 
spectral-domain tool able to probe the underlying phonon WSL.

Coherent phonon generation
is termed the impulsive generation of phonons using high power
ultrashort laser pulses.~\cite{coherent} In the case of THz 
vibrations, femtosecond pulses are required. To the best of our
knowledge the generation mechanism is still an open issue, and no
complete theory is available to describe the processes involved.
We briefly describe next a model for pump and probe coherent
phonon generation and detection based on a photoelastic coupling 
between light and acoustic phonons.~\cite{Mariano}
This mechanism is the only active when pump and probe are below 
the gap. Above the gap other mechanisms can contribute, but we expect 
the main conclusions to remain essentialy valid.
Any arbitrary time and position
dependent phonon displacement in the structure $w(z,t)$ can be expressed in 
terms of the phonon eigenstates as 
$w(z,t)=\int r_\omega u(z) sin(\omega t)d\omega$. 
Assuming that phonons are generated coherently through a 
photoelastic mechanism by a femtosecond pulse
(modelled as $E_0(z,t) \propto \delta(t)exp(ik_Lz)$), the coefficients 
for the above expansion can be obtained as~\cite{Mariano}
$r_\omega=\frac{1}{\omega}\int_0^L p(z) \left|E_0(z)\right|^2 
\frac{\partial u(z)}{\partial z} dz$. 
Here L is the length of the sample, $p(z)$ is the  photoelastic 
constant which is assumed constant in each layer, 
and ${\partial u(z)}/{\partial z}$ is the strain associated with 
an eigenstate of energy $\omega$.
We note that the FS ($q=0$) Raman cross section discussed above
is proportional to $\left|r_\omega\right|^2$.\cite{Trigo}
Once this excitation is generated, it evolves according to the
time dependence $sin(\omega t)$ and can be detected by a delayed,
lower power, probe pulse that senses the time dependent change of
reflectivity.~\cite{Thomsen} The change in reflectivity can be 
calculated as $\Delta r(t) \propto \int_0^\infty \Delta\epsilon(z,t)exp(2ikz)dz$, 
where the probe pulse has been assumed to be proportional to $e^{ikz}$.~\cite{Thomsen} 
Introducing again the photoelastic coefficient to relate $\Delta\epsilon$
with the strain ${\partial w(z,t)}/{\partial z}$, the time dependent 
change in reflectivity can be written as 
$\Delta r(t) \propto \int_0^\infty p(z)\frac{\partial w(z,t)}{\partial z} exp(2ikz)dz$.
If $p(z)$ is constant, this equation implies that only
phonons with $q=2k$ can be probed. On the other hand, the detector's 
$p(z)$ can be designed to access other excitations by backfolding
the phonon dispersion. Comparing this equation with the Raman
cross section,~\cite{Trigo} it is easy to see that the
observable phonon spectrum is related to the BS Raman scattering 
component of the detector structure.

In order to generate the quasi-monoenergetic phonons required
to induce BO, and to monitor the time evolution
of the latter, we have conceived a monolithic source-sample-detector
device (see the 
scheme in Fig.~\ref{fig4}). The first GaAs/AlAs SL (SL/s in Fig.~\ref{fig4}) 
acts as the phonon source and is designed to generate, by excitation with a fs 
impulsion above the gap, an elastic pulse with energy centered at 
18.3~cm$^{-1}$ and width equal to $\sim 0.8$~cm$^{-1}$.~\cite{coherent} 
It is made of 30 periods of $43.2\AA/51.5\AA$ GaAs/AlAs. The layer widths 
determine the energy of the generated pulse, while the number of
periods define, due to finite size effects, its width.
The coefficient $r_\omega$ of the phonon pulse (or equivalently 
the FS Raman spectra) generated in
this SL is shown in the inset on Fig.~\ref{fig4}. Once generated,
the coherent phonons propagate into the structure and act as the phonon
source $g(\omega)$ for exciting Bloch oscillations. Between the SL source 
and the structure, a GaAs 200nm buffer layer is introduced to screen
the residual pump power, not absorved in the SL, from impinging into 
the cavity structure and generating unwanted frequencies. 
The $g(\omega)$ phonon pulse, on the other hand, 
propagates through the GaAs layer basically unaltered. 
The cavity structure is identical to the one described above. 
Once within this structure,
Bloch oscillations develop and part of the energy is lost to the substrate 
upon each turn. Their effect at the right end side of the sample is 
probed by the second SL (SL/d in Fig.~\ref{fig4}), 
which acts as an energy selective detector of 
the Bloch oscillations through their effect in the time variation of 
the reflectivity.~\cite{Mariano} To keep this time variation simple,
a second GaAs layer is introduced to stop the probe beam from being 
modulated also by the cavity structure. The 
BS Raman spectrum of the SL defines the detector's bandwidth, which we have 
designed to include the pulse $g(\omega)$. It consists of a 20 periods 
$38.1\AA/68.1\AA$ GaAs/AlAs SL. Its BS spectrum, calculated for a 750~nm probe 
pulse, is shown in the inset on Fig.~\ref{fig4}.
The device is terminated by a thick ($> 1\mu$m) Ga$_{0.5}$Al$_{0.5}$As
layer which acts both as a stop layer for chemically etching the GaAs
substrate, and as a window for accessing the detector SL from the 
back. In Fig.~\ref{fig4} we present calculations of the reflectivity
change as a function of time. Fast oscillations dominate the reflectivity,
corresponding to the frequency of the coherently excited phonons 
(18.3 cm$^{-1})$. In addition, this fast component is amplitude modulated 
by an envelope whose frequency is determined precisely by the Bloch 
oscillation period. The Fourier transform of the time dependent reflectivity 
clearly shows the WSL frequencies within an envelope determined by
the input source $g(\omega)$ and the detector bandwidth.

In conclusion we have extended concepts previously discussed in
the context of electronic and photonic properties of solids to
acoustic phonon physics. We have described semiconductor 
structures based in recently reported acoustic cavities and 
achievable with actual growth technologies, displaying phononic
Stark ladders and capable of sustaining Bloch oscillations. 
We have also shown how Raman scattering and coherent phonon 
generation provide spectral and time domain probes, respectively,
of these acoustic phenomena. Structures as the one described
here can be exploited to enhance the coupling between sound
and other excitations (electrons and photons), and as the feedback
high-Q resonator of a phonon laser. Moreover, engineered phonon
potentials are not limited to linear dependencies that mimic
electric fields but can take arbitrary shapes. This opens the
way to novel phonon devices based in the discrete confined
states of acoustic cavities.

We acknowledge M. Trigo and B. Perrin for enlightening discussions 
and G. Malpuech for useful information concerning the porous silicon 
optical Bloch structures. AF also acknowledges support from the 
ONR (US).

\newpage
\begin{figure}
\caption{(a) Phonon band 
structure. The central white region limited by the thicker black 
forbidden bands corresponds to the
cavity-mode miniband. The white and thinner black regions above and below this
bands arise from Bragg oscillations. The arrows indicate the energy 
of the pulses displayed in Fig.~2. (b) Phonon reflectivity. In (a) and (b) shaded 
areas indicate the WSL region. (c) Phonon displacement as a function of position 
and energy.  Darker regions indicate larger acoustic phonon intensities. The arrows
label the WSL energies. The solid lines limit the ``forbidden'' bands. 
See text for details.           
\label{fig1}}
\end{figure}

\begin{figure} [!h]                                                                   
\caption{$log(u(z,t))$ for gaussian phonon wavepackets
centered at the two energies indicated with arrows in 
Fig.~\ref{fig1}(a), and FWHM given by $2\sigma=1$~cm$^{-1}$.
\label{fig2}}                               
\end{figure}                                                                    

\begin{figure} [!h]     
\caption{Back and forward 550~nm Raman
scattering spectra corresponding to the acoustic cavity structure
with $N_{c}=25$ described in the text.
For comparison the corresponding phonon reflectivity 
is also shown. The inset highlights the Stark ladder 
spectral region (17-19.5~cm$^{-1}$).
\label{fig3}}                               
\end{figure}

\begin{figure} [!h]                                                                  
\caption{Top-left: Source and detector SL's Raman spectra. Top-right:
Scheme of the proposed monolithic source-sample-detector phonon device
(see text for details). Bottom-left: Time variation of the reflectivity 
probed at the detector SL. Bottom-right: Fourier transform of the 
detected reflectivity change.
\label{fig4}}                               
\end{figure}


\newpage                                                                        
\begin{references}    

\bibitem{Bloch} F. Bloch, Z. Phys. {\bf 52}, 555 (1928); 
C. Zener, Proc. R. Soc. London A {\bf 145}, 523 (1934).


\bibitem{SL's} See, for example, J. Feldman {et al.}, 
Phys. Rev. B {\bf 46}, 7252 (1992); C. Waschke {\it et al.},
Phys. Rev. Lett. {\bf 70}, 3319 (1993).

\bibitem{Malpuech} G. Malpuech and A. Kavokin,
Semicond. Sci. Technol. {\bf 16}, R1 (2001),
and references therein.

\bibitem{PRL's} R. Sapienza {\it et al.}, Phys. Rev. Lett. {\bf 91},
263902 (2003); V. Agarwal {\it et al.}, Phys. Rev. Lett. {\bf 92},
097401 (2004).

\bibitem{Trigo} M. Trigo {\it et al.}, Phys. Rev. Lett. {\bf 89}, 
227402 (2002); see also J. M. Worlock and M. L. Roukes, 
Nature {\bf 421}, 802 (2003).

\bibitem{Philosophical} W. Chen {\it et al.}, Philosophical Magazine B {\bf 70}, 687 ().

\bibitem{Narayanamurti} V. Narayanamurti {\it et al.}, 
Phys. Rev. Lett. {\bf 43}, 2012 (1979).

\bibitem{Lacharmoise} P. Lacharmoise {\it et al.}, Appl. Phys. Lett.
{\bf 84}, 3274 (2004).

\bibitem{Jusserand} B. Jusserand and M. Cardona, in {\em Light Scattering in Solids V}, 
edited by M. Cardona and G. G\"untherodt                 
(Springer, Heidelberg, 1989), p. 49.

\bibitem{Pinan} A two meter SOPRA double grating spectrograph can have a
resolution around 0.02~cm$^{-1}$. A tandem Fabry-Perot Raman monochromator 
set-up can improve this perfomance down to 0.005~cm$^{-1}$. See, e.g.,
J. P. Pinan {\em et al.}, J. Chem. Phys. {\bf 109}, 5469 (1998).

\bibitem{coherent} T. Dekorsy, G. C. Cho, and H. Kurz, 
in {\em Light Scattering in Solids VIII}, 
edited by M. Cardona and G. G\"untherodt                 
(Springer, Heidelberg, 2000), p. 169.

\bibitem{Mariano} M. Trigo, T. Eckhause, and R. Merlin, private communication. 

\bibitem{Thomsen} C. Thomsen  {\it et al.}, Phys. Rev. B {\bf 34},
4129 (1986).



\end{references}
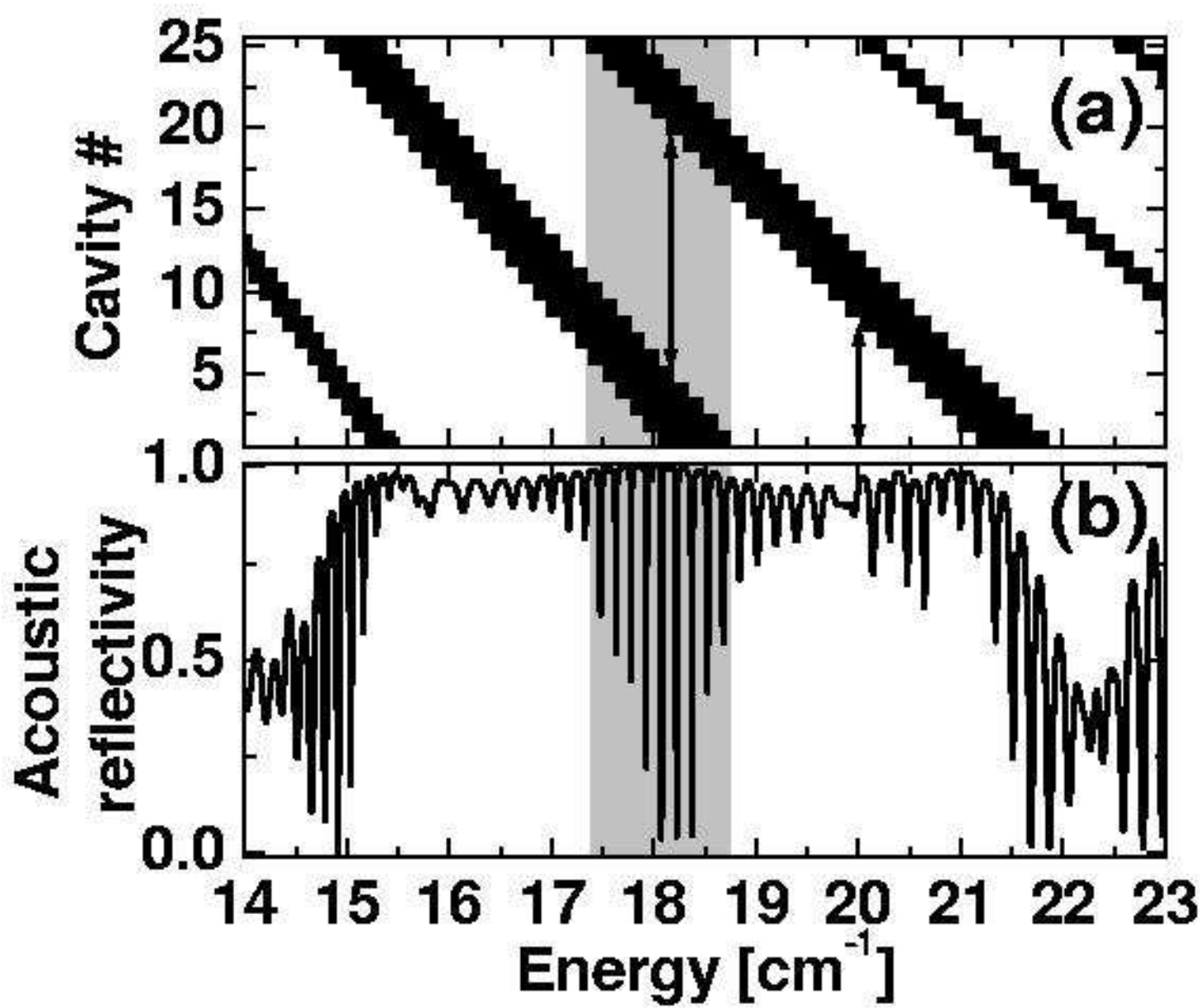
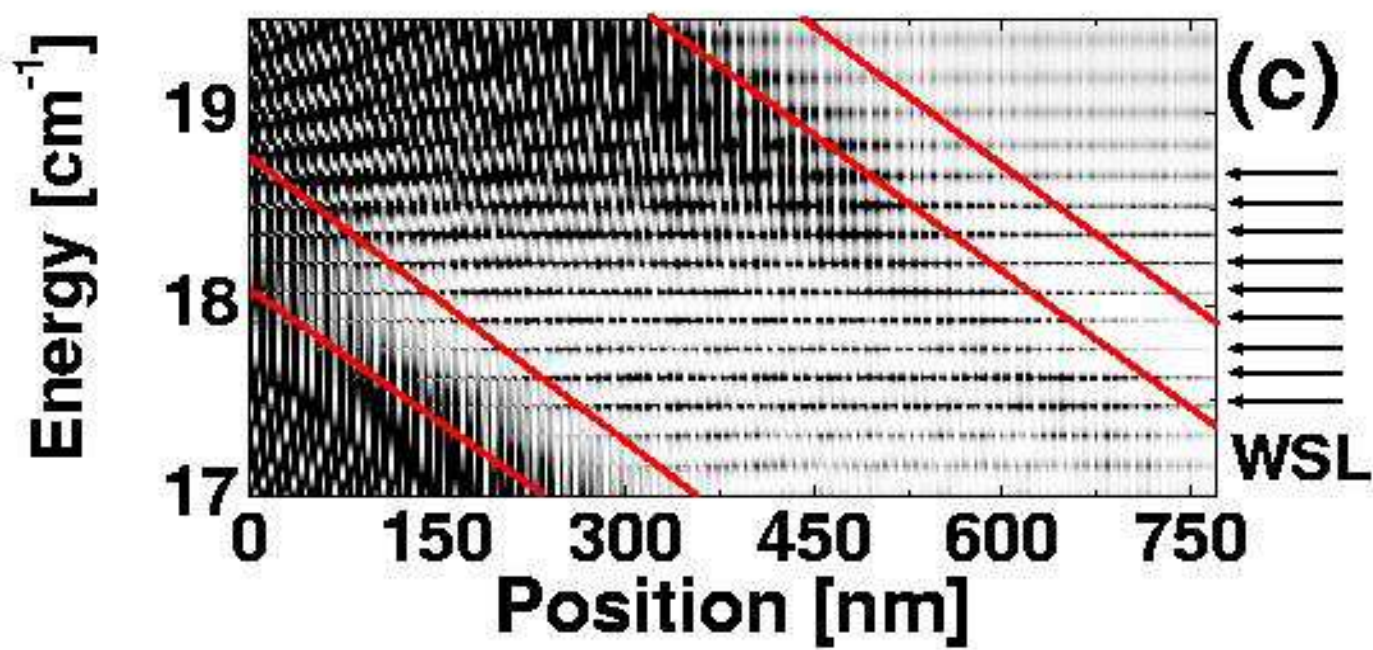
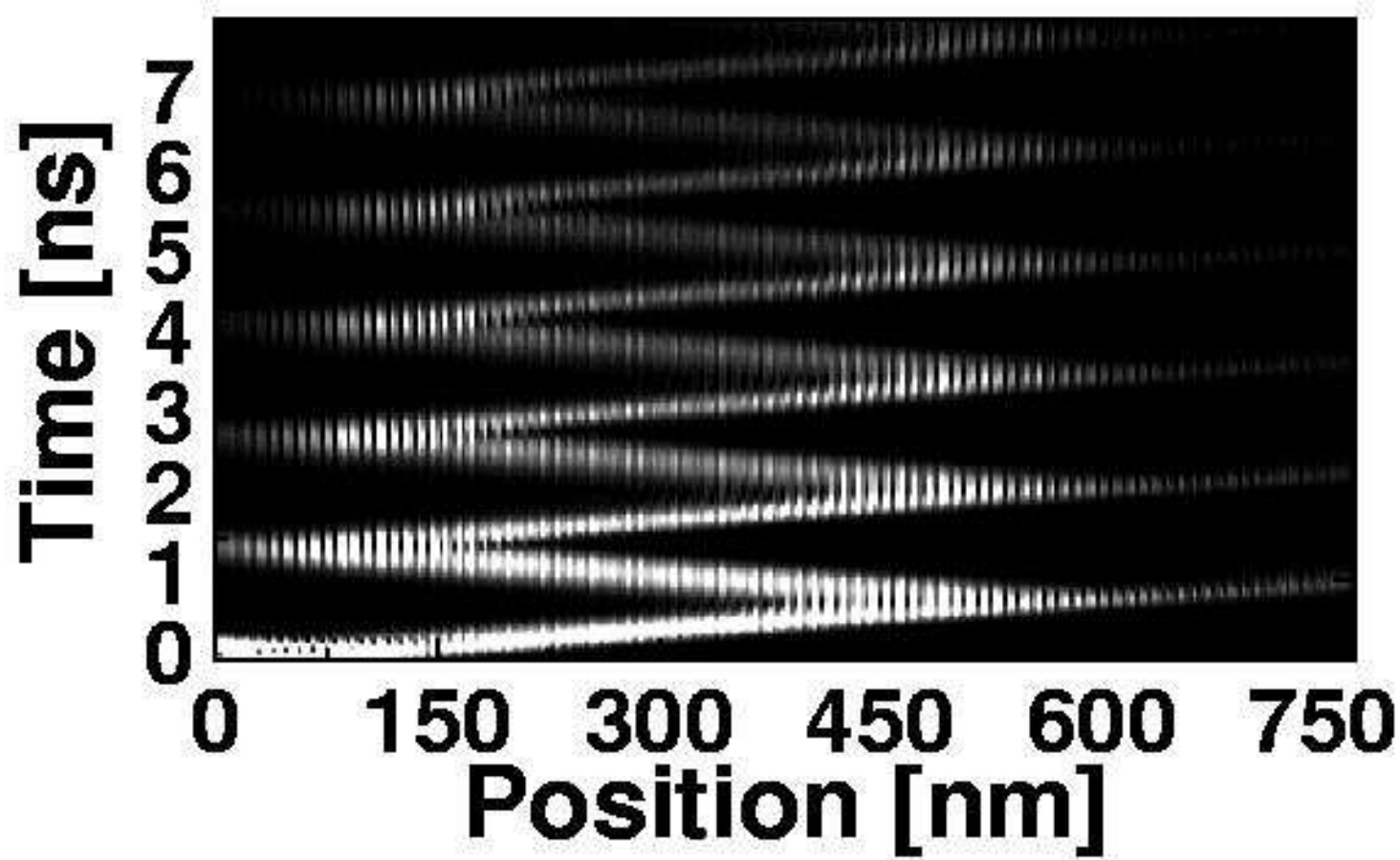
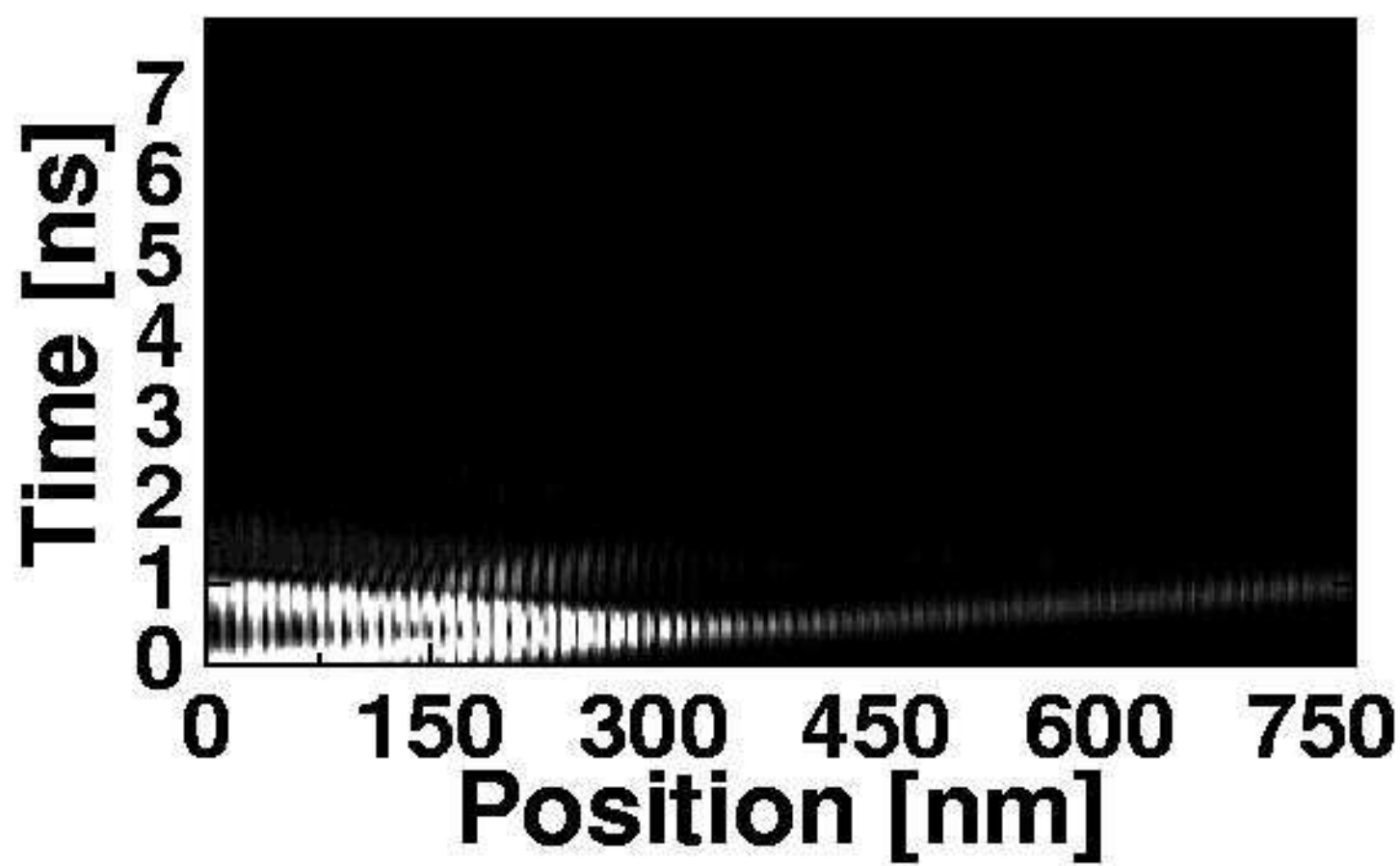
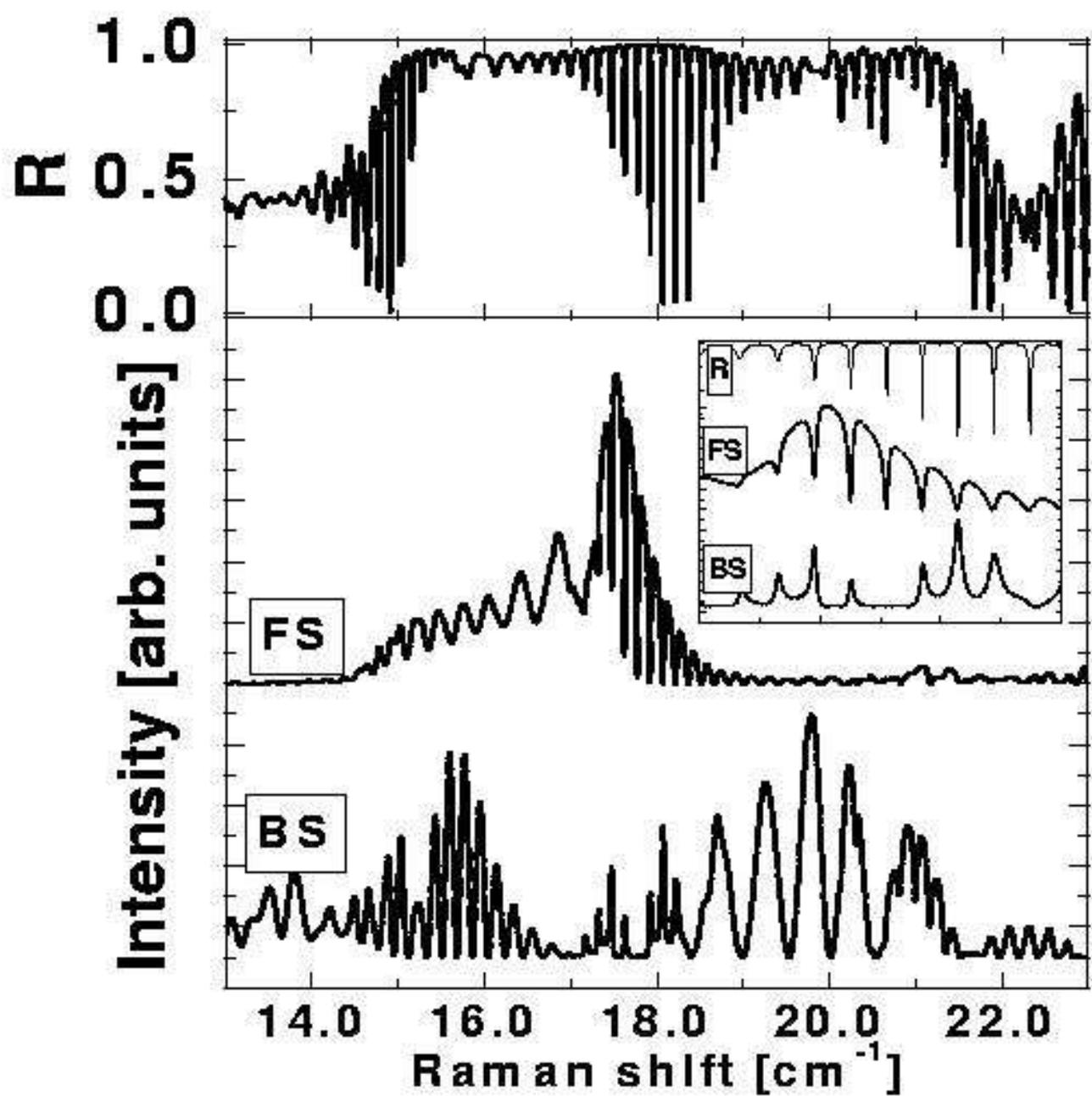
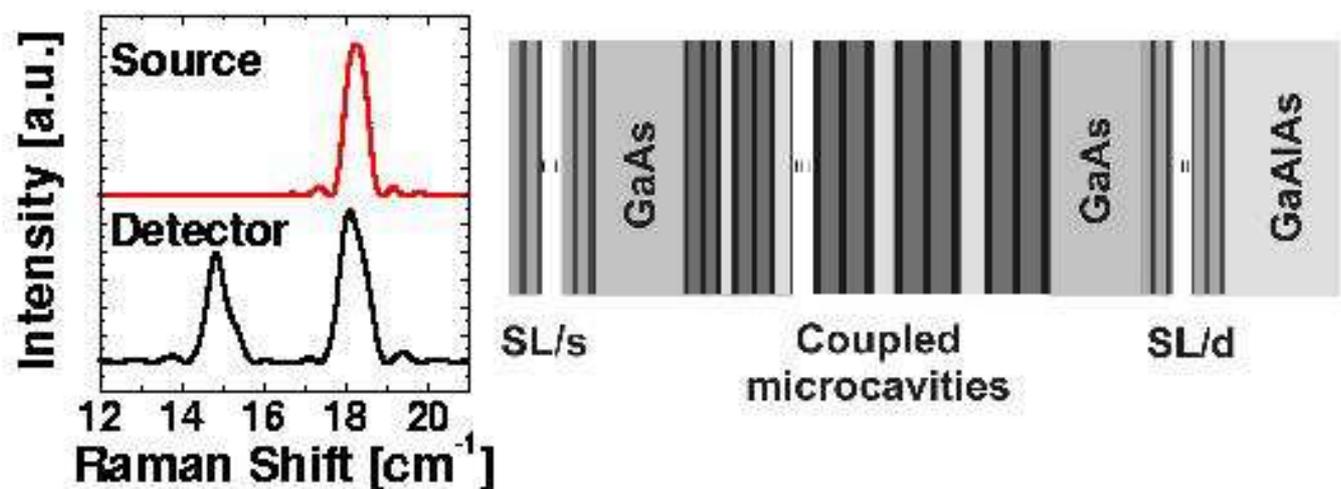
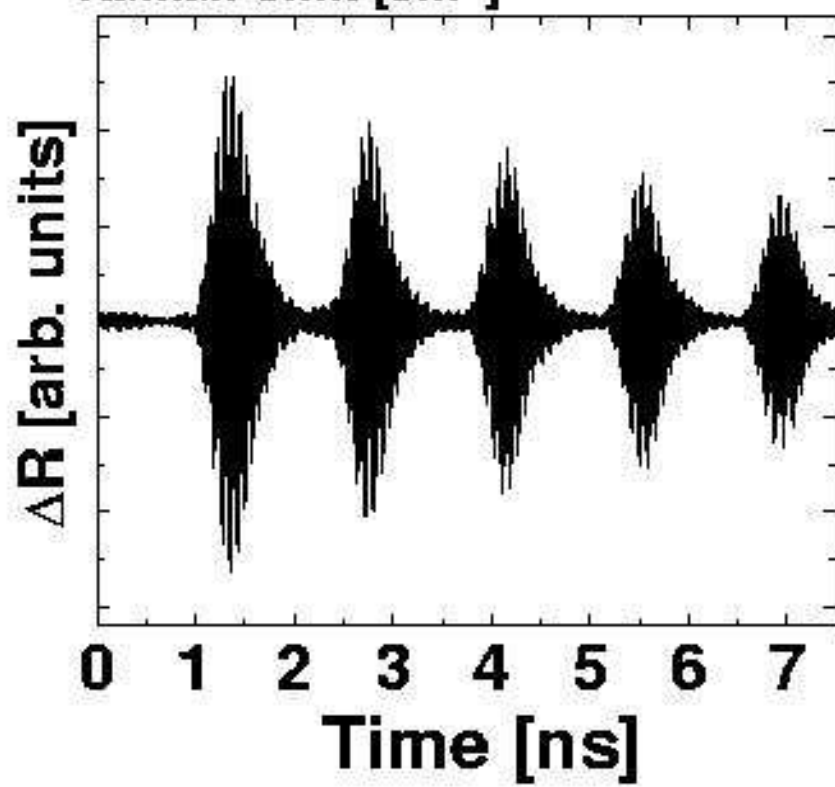
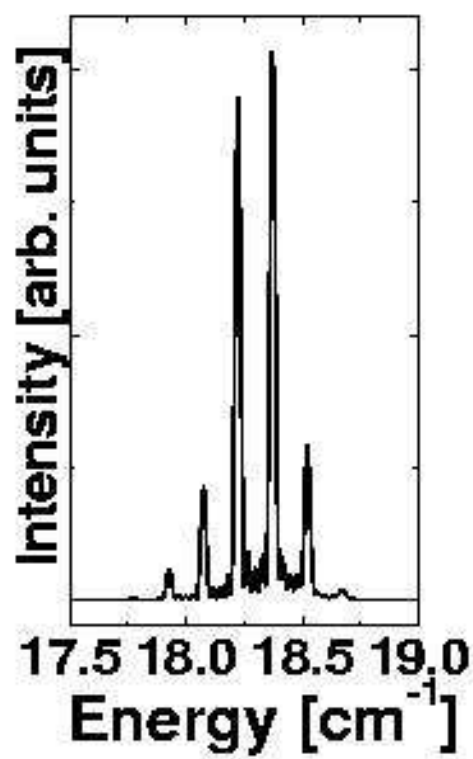
\end{document}